# Design of a large-scale superconducting dipole magnet for the CEE spectrometer

Yuquan Chen[a,c], Wei You[a,c], Jiaqi Lu[a,c], Yujin Tong[a], Luncai Zhou[a], Beimin Wu[a,c], Enming Mei[a,c], Wentian Feng[a,b], Xianjin Ou[a], Wei Wu[c,d]*, Qinggao Yao[a,b], Peng Yang[a], Yuhong Yu[a,b], Zhiyu Sun[a,b]

[a]Institute of Modern Physics, Chinese Academy of Science: No.509 Nanchang Road, Lanzhou, Gansu, 730000
[b]University of Chinese Academy of Sciences: No. 1 Yanqihu East Road, Huairou District, Beijing, 101408
[c]Advanced Energy Science and Technology Guangdong Laboratory: No. 29 Sanxin North Road, Huizhou, Guangdong, 516000
[d]Shanghai APACTRON Particle Equipment Co., Ltd: No. 1180 Xingxian Road, Jiading District, Shanghai, 201815
E-mail address: wuwei@impcas.ac.cn.

The CSR External-target Experiment (CEE) is a large-scale spectrometer under construction at the Heavy Ion Research Facility in Lanzhou (HIRFL) for studying the phase structure of nuclear matter at high baryon density and the equation of states of nuclear matter at supra-saturation densities. One of the key components is a large acceptance dipole magnet with a central field of 0.5 T and the homogeneity of 5% within a 1 m long, 1.2 m wide, and 0.9 m high aperture. Detectors will be installed within this aperture. An innovative design for the superconducting detector magnet is proposed that goes beyond the conventional approach. The magnet is designed as a coil-dominant type, with conductors discretized on a racetrack-shaped cross-section to generate the necessary fields. A warm iron yoke is used to enhance the central field and minimize the stray field. The magnet has overall dimensions of 3.4 meters in length, 2.7 meters in height, and 4.3 meters in width. The coils will be wound using a 19-strand rope cable comprised of 12 NbTi superconducting wires and 7 copper wires. The ratio of copper to superconductor of the cable is 6.9. The keel supports serve as the primary structural support for the coils to withstand the electromagnetic force. The coils will be indirectly cooled by liquid helium within three external helium vessels. To ensure reliable protection of the magnet during a quench, an active protection method combined with quench-back effect is employed. In this paper, we mainly present the detailed design of the magnetic field, structure, quench protection and cryostat for the spectrometer magnet.

1. **Introduction**

   The CEE spectrometer is under construction at the Institute of Modern Physics (IMP) in China. It was proposed to study the density dependence of nuclear symmetry energy, the EOS at supra-saturation density and the rich Quantum Chromo Dynamics (QCD) phase at high-density and low-temperature [1,2]. The CEE system mainly consists of a micro-pixel beam monitor, a large volume Time Projection Chamber (TPC), a Multi-Wire Drift Chamber (MWDC) array, Time-of-Flight detectors (TOF) and a large acceptance superconducting dipole magnet. Fig. 1 presents the layout of the CEE spectrometer [3]. The large TPC will be installed at the center of the dipole magnet to track the trajectories of the secondary particles. Therefore, the aperture of the dipole magnet needs to be sufficiently large to accommodate the detectors.

   According to the requirement, the free aperture should be no less than 1.6 m for the vertical gap and 1.6 m for the horizontal space. The magnet is designed with a racetrack-shaped cross section, which provides greater cost-effectiveness compared to a circular shape when considering field homogeneity. The winding has a saddle-shaped coil end. To optimize the aperture field, the conductors are discretized along the azimuthal direction on the arc side of the racetrack-shaped cross section while a few conductor turns are positioned on the straight side to regulate the field. Each pole of the magnet is divided into three coils. To minimize coil inductance and simplify the winding process, rope-type superconducting cables are used and embedded into pre-machined channels. The copper to superconductor ratio of the cable is 6.9. The operating current at 0.5 T is 1580 A.

   The deformation of the coil structure induced by thermal shrinkage will be significant due to its large size. Meanwhile the electromagnetic force on the coil is also significant which can lead to substantial expansive deformation. Based on the distribution characteristics of the electromagnetic force, a robust support structure made of stainless steel is introduced to resist deformations. The deformations and the stresses of the coil and support structure under various load conditions are calculated by finite element method.

For the large-scale coil with 3 m long, 1 m high and 3.5 m wide, it is inefficient and uneconomical to cool it down by helium bath with a very large and sophisticated cryogenic vessel. Therefore, an indirect cooling method will be employed [4,5]. Three racetrack type helium vessels are installed at both ends and middle section of the coil to provide cold source. Additionally, the straight sections of the coils will be cooled by liquid helium pipes soldered outside the coil cases. The coil case consists of a bottom plate, two side plates and a cover plate. The bottom plate and side plate are made of stainless steel while the cover plate is made of aluminum alloy.

The stored energy of the dipole magnet is approximately 3 MJ at the nominal current of 1580 A. Protecting the magnet during a quench with such a large stored energy is a significant challenge. The quench propagation velocity in the transverse direction is crucial for the quench protection. In order to speed up the quench propagation, cooper wires are wound together with the cables in the channels to act as heaters and trigger the quench-back. The quench behavior is simulated with quench-back and also calculated without quench-back in case it is invalid.

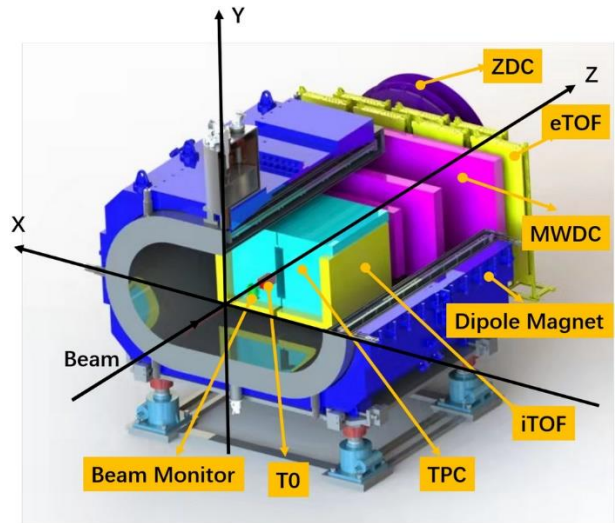

Fig. 1. The layout of the CEE spectrometer

2. **Magnetic field design**

According to the magnetic field requirements outlined in Table 1, the CEE dipole magnet will adopt a coil-dominated design with NbTi superconducting coil and warm iron yoke which is so called superferric magnet. Due to severval large detectors such as TPC and iTOF will be positioned within the aperture of the magnet, adequate space must be provided to accommodate them. The available space should be at least 1.6 m in height and over 1.6 m in width. The dipole magnet features two good field regions, each of which is 0.5 m wide, 1 m long and 0.9 m high. The interval of the two regions is 0.2 m. The magnetic field homogeneity in these regions is within 5%.

The coil design is based on the concept of the discrete current distributed on the racetrack shell whose geometric shape can provide a cost-effective large magnetic field region. To mitigate risks, the coils of each pole are divided into three blocks. Two saddle-shaped coil blocks are positioned at the arc segments of the racetrack shell, and one flat-shaped coil block is positioned at the straight part. The six individual coils are connected in series.

**Table 1**
The physical requirement of the CEE dipole magnet

| | |
|---|---|
| Central field | 5000 Gauss |
| Good field region (two regions) | 0.5 m wide, 1 m long and 0.9 m high with an interval of 0.2 m |
| Homogeneity in the good field regions | < 5% |
| Aperture height | 1.6 m |

The 3D magnetic analysis and optimization of the dipole magnet are performed by the CST Studio Suite. Fig. 2 depicted the calculated magnetic field model. The model consists of the iron yoke, coils, good field regions and air background. Each pole has three coils, referred to as inner coil, middle coil and outer coil based on their positions. The middle coil and outer coil are saddle-shaped, while the inner coil is flat-shaped. The central magnetic field is mainly produced by the middle and outer coils on the arc side of the racetrack shell while the inner coil on the straight segment is used to regulate the magnetic field. The iron yoke at ambient temperature is positioned outside the coils to enhance the central magnetic field and minimize the fringe field. The weight of the iron yoke is 68 ton. The maximum and minimum magnetic field in the good field regions are 5255 Gauss and 5010 Gauss respectively, meeting the homogeneity requirement of 5% (see Fig. 3). The maximum field is situated at the upper and lower surfaces which are close to the iron yoke.

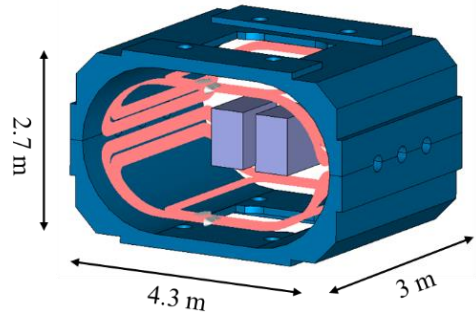

Fig. 2. The electromagnetic model created by the CST software

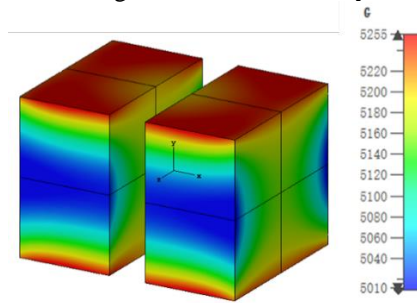

Fig. 3. Distribution of magnetic field on the good field regions

The main design parameters are listed in Table 2. The 19-strands rope cable is used for the coil winding. For operational stability, the copper to superconductor ratio of the cable is designed to be 6.9. The cable is composed of 12 superconducting wires and 7 copper wires. The superconducting wire and copper wire both have a diameter of 0.6 mm. The maximum field on the coils is approximately 1.9 T, with a load line margin of 26.5% at 4.2 K. The current sharing temperature is 5.3 K, with a temperature margin of 1.1 K.

**Table 2**
The main design parameters

| | |
|---|---|
| Operating current | 1580 A |
| Dimensions of the coil | 3.1 m (length)×3.54 m (width) |
| Inductance at nominal current | 2.3 H |
| Stored energy | ~3 MJ |
| Maximum magnetic field on the coil | 1.9 T |
| Number of turns per coil | 132/136 |
| Io/Ic ratio along the load line | 73.5 % |
| Iron yoke weight | 68 ton |
| Cable type | 12 SC wires + 7 copper wires |
| Copper : SC ratio of cable | 6.9 |
| Diameter of insulated cable | 3.2 mm |
| Diameter of SC wire | 0.6 mm |
| Copper : SC ratio of wire | 4 |
| Critical current @4.2K | 280A @ 2T |
| | 160A @ 4T |

The saddle-shaped coil model is shown in Fig. 4. For easy winding, U-shaped grooves are machined onto the coil former [6,7]. The interval between two adjacent grooves is 1 mm. To minimize the total length of the coil, close winding is applied at the coil end. Two layers of cables are wound into one channel on one former. Each coil consists of two layers of formers made of G10 material. Fig. 5 displays the cross-section of two layers of G10 formers with grooves and the 19-strand ropes on the straight part of the winding. Copper strips are co-wound into grooves above the cables, which is designed to act as a heater to accelerate the expansion of the normal zone during quench.

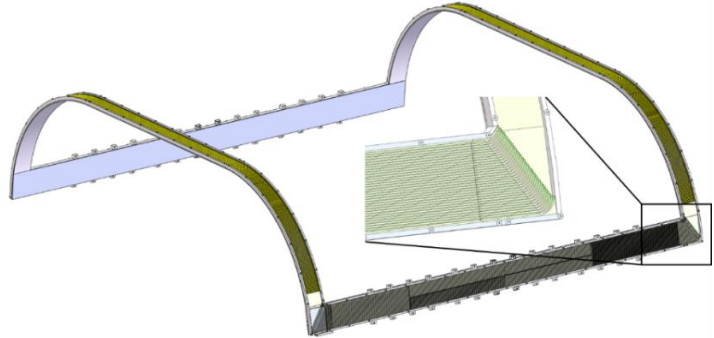

Fig. 4. The saddle coil model with wire grooves and coil case

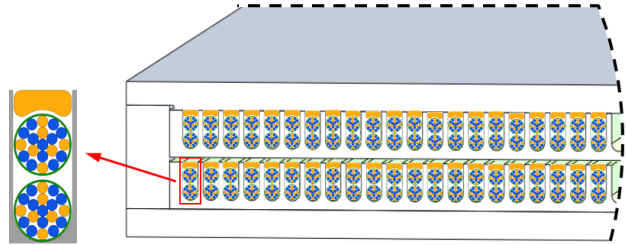

Fig. 5. Cross section of two layers G10 formers with the groove and the 19-strands ropes

3. **Mechanical design**

The cold mass of the CEE magnet consists of six coil blocks, support frameworks and three liquid helium vessels as illustrated in Fig. 6. The helium vessels are located outside the coils which will be of benefit to constraint the expansion of the coils during energization. Moreover, the thermal contact between the coils and the helium vessels will be improved due to the contraction of the vessels and expansion of the coils. The straight sections of the coils are secured on the support framework with bolts, while the coil ends are connected to the helium vessels using fixtures.

Considering the significant deformation of the cold mass caused by thermal contraction during cooling down, both rods and post supports are utilized to sustain it. The weight of the cold mass is about 5 ton. The fixation system of the cold mass comprises seven post supports assembled at the bottom and four rods installed at the top. Additionally, three rods are applied on each side of the straight section to resist against the electromagnetic force.

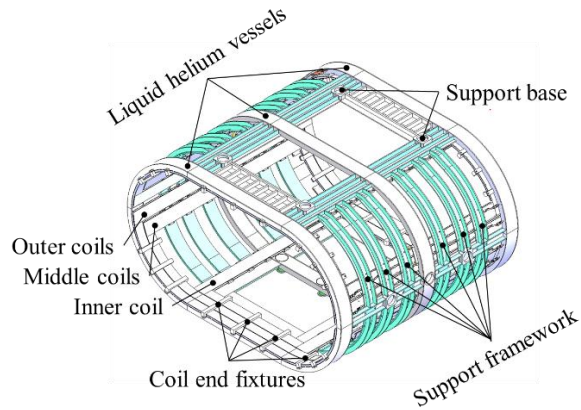

Fig. 6. Structure model of the cold mass including rods and post supports

Finite element models are developed using ANSYS software to assess the displacements and stresses of the cold mass under the loads of gravity, thermal shrinking during cool-down, and Lorentz forces after energization. The model also takes the cryostat into account for the application of boundary condition. The lower surface of the vacuum vessel of the cryostat is fixed which is assembled on the iron yoke. Furthermore, contact pairs are established between the post supports and the vacuum vessel. The contact type between the middle post support and the vacuum vessel is defined as bonded, while the other post supports are defined as no separation. The rods are unrestricted during cool-down but fixed during energizing. The boundary conditions of the cold mass model are shown in Fig. 7.

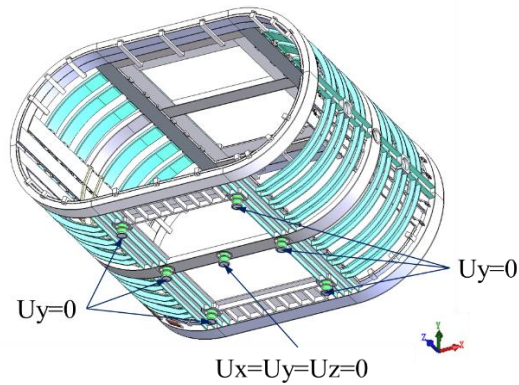

Fig. 7. Boundary conditions of the model on the post supports.

Under the effect of gravity, the maximum displacement of the cold mass is 0.3 mm which is mainly along the vertical direction. Fig. 8 shows the distributions of the total deformation and the Von Mises stress of the cold mass. The maximum Von Mises stress of the cold mass is about 20 MPa which is at a low level.

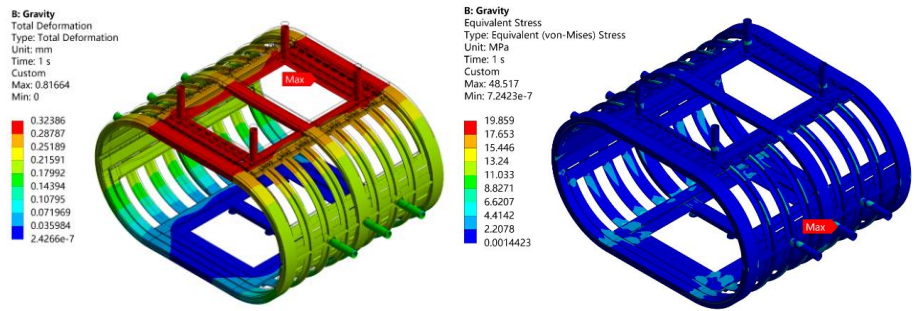

Fig. 8. Displacement (left) and Von Mises stress (right) of the cold mass under gravitational loads.

During the cool-down process, the cold mass shrinks towards the fixed bottom center position. The total displacement of the cold mass after cool down is shown in Fig. 9. The maximum displacements in the length, width and height directions are respectively 5.3 mm, 6.8 mm and 1.6 mm. The thermal contraction of the cold mass in the length direction is about 0.35%. Thermal stress will be generated due to the inconsistent shrinkage coefficients of each component. Fig. 10 shows the Von Mises stress on the cold mass and coils after cool down. The maximum stress is concentrated in some very small local areas which is not a real value and caused by the binding of the contacted surface. But the stress is accurate on the areas away from these regions. The Von Mises stress on the coils is below 150 MPa and on the helium vessels, it is below 250 MPa.

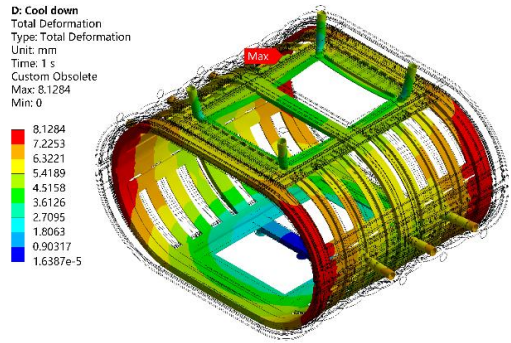

Fig. 9. Total displacement of the cold mass after cool down

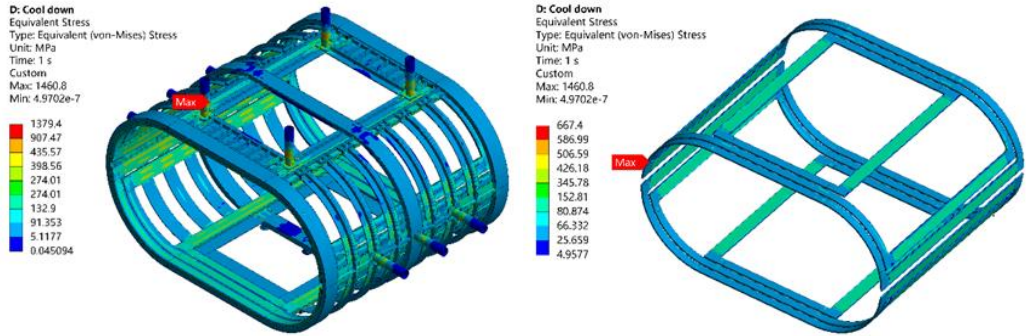

Fig. 10. Von Mises stress on the cold mass (left) and coils (right) after cool down

The displacements and stresses are calculated when the magnet is energized, taking into account the effects of gravity and cooling down. Fig. 11 shows the displacement of the cold mass after energization. The electromagnetic force expands outward in the width direction which is opposite to thermal contraction. However thermal contraction displacement remains the major component of the total displacement. Fig. 12 shows the Von Mises stress on the cold mass and coils after energization.

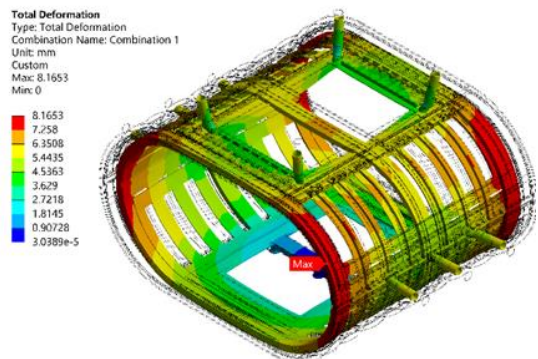

Fig. 11. Displacement of the cold mass after energization

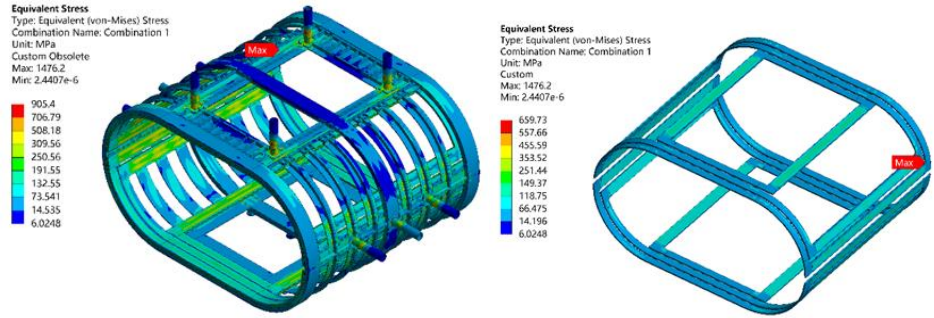

Fig. 12. Von Mises stress on the cold mass and coils after energization

## 4. Quench protection

For such a superconducting magnet having a large size and huge stored energy, an active protection scheme is preferred to protect the CEE dipole magnet during the quenching events. Two approaches that improve the quench propagation velocity are employed to quickly transfer voluminous parts of the coil to the normal-conducting state so that the hot-spot temperature and the peak voltage-to-ground can be limited to an allowable range. The first method is using quench heaters which are made of flat copper wires and co-wound with the superconducting cables in the grooves. The quench heaters are powered by inductive current caused by the mutual inductance effect during energy discharge which is also a quench-back effect [8]. The co-wound copper wires are used for both quench detection and quench heating. The second method is the use of aluminum alloy as coil former which can also bring about a quench-back due to the heat generated by the eddy current during the magnet discharge.

Fig. 13 shows the schematic circuit diagram of the quench protection scheme, which consists of two circuits during the energy discharge. In the superconducting coil circuit, $R_d$ is the dumping resistor connected in parallel with the six superconducting coils. In the Cu coil circuit, $R_a$ is the adjustable resistor and D2 is diode module connected in series with the copper coils. In addition, the $R_a$ is used to control the induced current and voltage in the Cu coil and the D2 is utilized to blocking the induced current in the heater wire during the energy exciting process. Two set of insulated gate bipolar transistor (IGBT) switches were integrated with the power supply (PS) and the energy dumping system, which will be open once the normal zone is detected in the magnet. The quench is detected by measuring the voltage signals of entire SC coil and the copper wires. According to a certain proportion, the inductive voltage can be eliminated by subtracting the voltage signals of the copper coil from the SC coil, so as to extract the components of resistance voltage when the magnet is quenched [9].

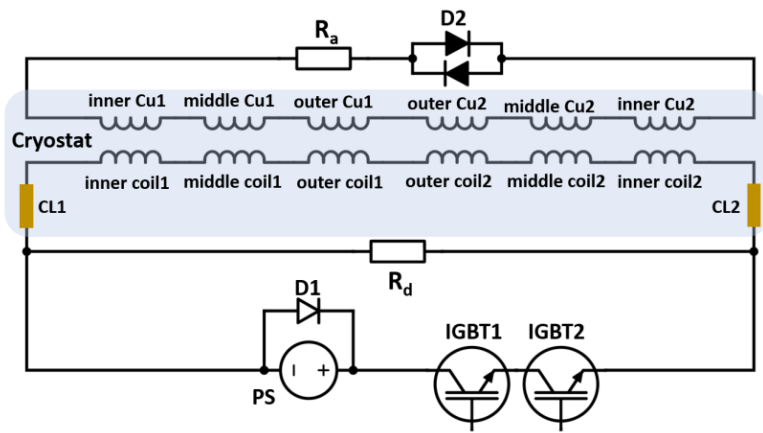

Fig. 13. the schematic circuit diagram of the quench protection scheme with the quench-back triggered by the co-wound heater wire

The quench transients are calculated using a two-dimensional (2D) electro-thermal coupling quench model [10] considering the quench-back effect triggered by the co-wound copper heater wire. As shown in Fig. 14, the quench-back behavior occur at about 0.5 s after the energy extraction starts, and the maximum hot-spot temperature reaches about 107 K. Meanwhile, the maximum voltage of the superconducting to ground is 474V due to the energy transfer to the dumping resistor ($R_d$=0.3Ω) after the IGBT opens.

In addition, Considering the conservative design to be self-protecting, the quench simulation without quench-back and dump resistor is also performed using 3D electro-thermal coupling quench model. Fig.15 compares the evolutions of the current and hot-spot temperature of the superconducting magnet for the case with quench-back and dumping resistor, only with dumping resistor and without quench-back and dumping resistor (i.e. the self-protecting case). As shown in Fig. 15, the hot-spot temperature of the magnet will reach about 180 K for the case of only using the dumping resistor and about 240 K for the case of self-protection after the energy discharge. Thus, the magnet could be also protected without the quench-back and dumping resistor and a lower peak hot-spot temperature of 107 K can be achieved when the quench detecting system and the quench-back work properly. Meanwhile, part of the stored energy can be brought out from the cryostat utilizing the active quench protection system, to decrease the pressure of the cryostat.

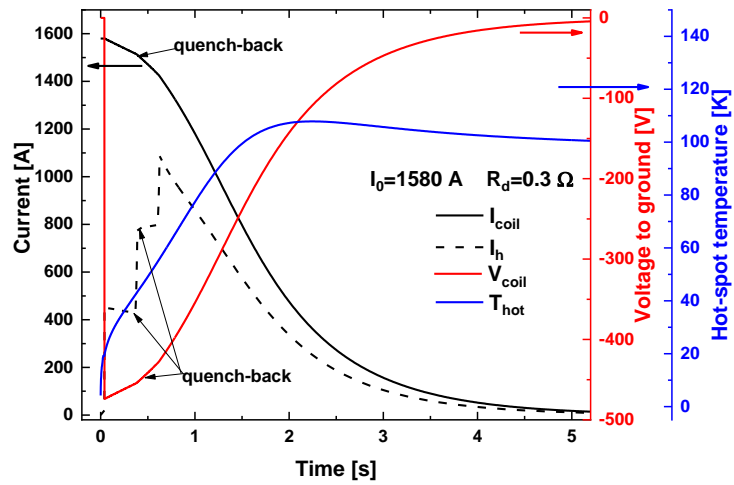

Fig. 14. Evolutions of the current, voltage and hot-spot temperature of the superconducting coil during the energy discharge process.

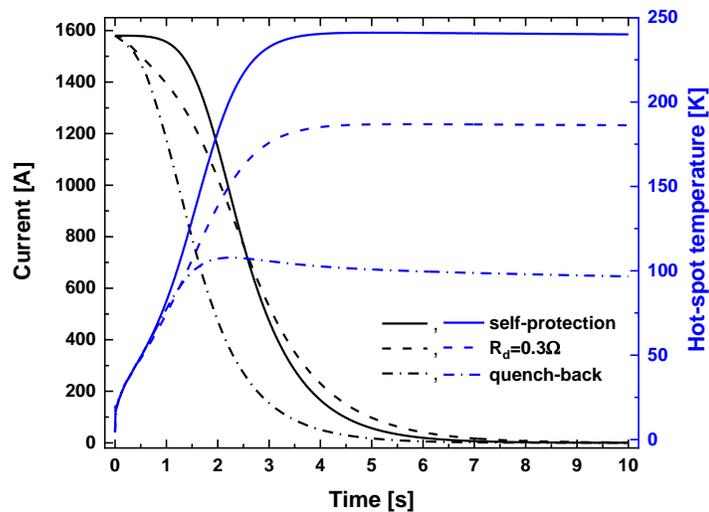

Fig. 15. Comparisons of the current and hot-spot temperature for the different quench protection scheme.

## 5. Cryostat design

The cryostat of the CEE magnet mainly consists of three liquid helium vessels, thermal radiation shields, superinsulation and the outer vacuum vessel, providing the cryogenic environment for the operation of the superconducting coils (see Fig.16). As mentioned above, the cold mass is quite large. Using a liquid helium bath for coil cooling would require a huge and bulky helium vessel, which is not cost-effective. Therefore, the indirect conduction cooling method is preferred rather than liquid helium bath [10]. Three individual helium vessels interconnected by pipes, serve as the cold source for conduction cooling. The helium vessels will be filled with approximately 234 L of liquid helium. Copper strips are also utilized to enhance the heat transfer between the cold source and the coils. Meanwhile, four two-stage GM cryocoolers are installed in the feed box to re-condense the evaporated helium, while two single-stage GM cryocoolers are employed to independently cool the thermal shield at bottom. The thermal shields are made of 1100 aluminum.

Herein, we describe details of the estimation of heat loads, thermal analysis and suspension structure analysis.

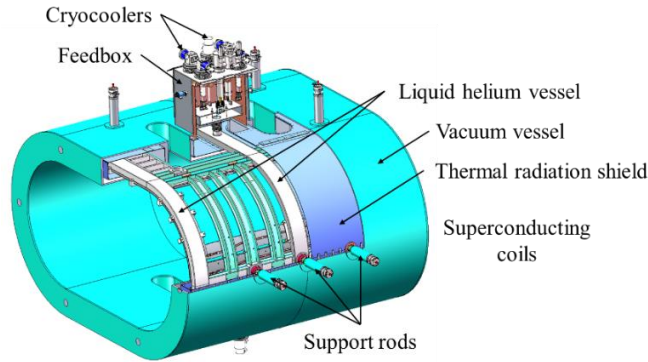

Fig. 16. Diagram of the cryostat design

5.1 Estimation of heat load

The heat load includes two parts: one is $1^{st}$ stage heat load at 50 K which comes from conduction heat through copper current leads and supports, ohmic heat from copper current leads and thermal radiation from RT to shield, the other one is $2^{nd}$ stage heat load at 4.2 K which comes from conduction heat through HTS leads and supports and thermal radiation from shield to cold mass [11]. Two pairs of 1000 A HTS current leads are utilized for the coils with operating current of 1580 A. Table 3 present the calculated heat load. The thermal shield consists of 80 layers of Multi-Layer Insulation (MLI).

**Table 3**
Estimates of heat load

| Location of the heat load | Heat load (W) |
|---|---|
| Conduction heat through copper current leads (×4) | 131.6 |
| Conduction heat through rods (×10) | 45 |
| Conduction heat through post supports (×7) | 35 |
| Thermal radiation from RT to shield (50K) | 37.5 |
| Total 1st stages load | 249.1 |
| Conduction heat through HTS current leads (×4) | 0.57 |
| Conduction heat through rods (×10) | 0.47 |
| Conduction heat through post supports (×7) | 0.35 |
| Joule heat from joints (×7) | 0.017 |
| Thermal radiation from shield to magnet | 0.5 |
| Total 2nd stages load | 1.9 |

5.2 Thermal analysis

The HTS current leads and thermal shield are conduction-cooled by copper plates and strips attached to cold heads of GM cryocoolers. The cold mass is indirectly cooled by both the liquid helium vessels and $2^{nd}$-stage cold heads. Fig. 17 shows the simulated temperature distributions of thermal shields by FM analysis. The conduction heat loads are applied to the rods and post supports, while the thermal radiation heat load is applied to the surface of the thermal shield. The maximum temperature difference on the thermal shields is about 14 K. The maximum temperature is located at the central position of the outer layer, which is the farthest point from the cold source.

The cold mass is cooled by heat conduction through the helium vessels and cryocoolers. Copper plates and strips are used to connect the cold source and cold mass as shown in Fig. 18. Fig. 19 shows the calculated temperature distribution on the cold mass. The maximum temperature is located at the support seats. Fig. 20 shows the temperature distribution on the coils. The temperature difference is within 0.1 K, which is below the share temperature of 5.6 K.

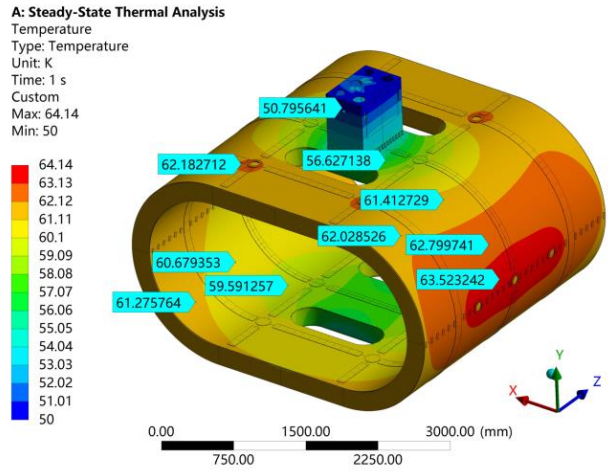

Fig. 17. Thermal analysis of the thermal shields

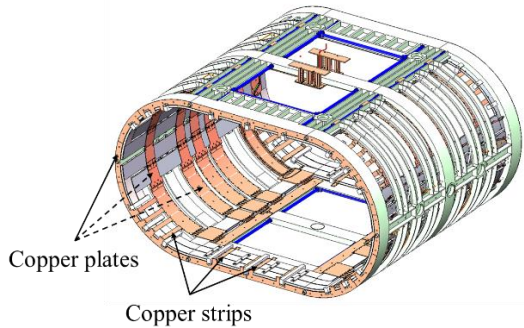

Fig. 18. Structure of the cold mass with heat conduction component

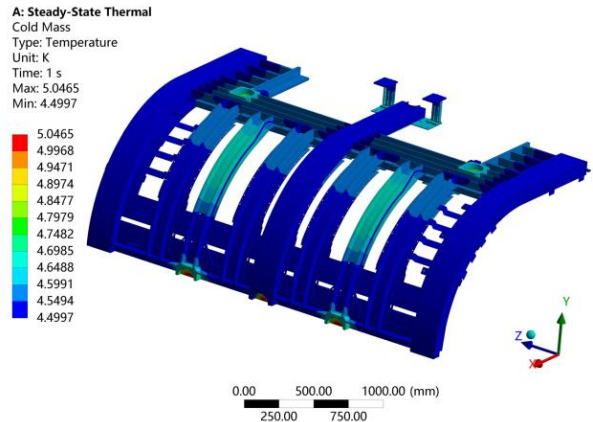

Fig. 19. Thermal model of the cold mass including coils, helium vessel and support framework

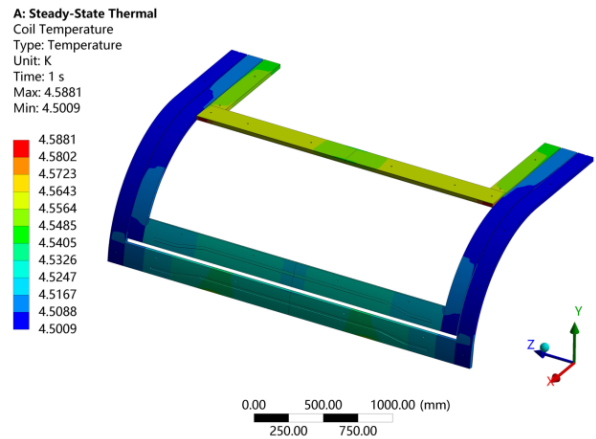

Fig. 20. Temperature distribution on the coils

## 6. Summary

A large-scale superconducting dipole magnet has been designed for the CEE spectrometer including magnetic field calculation, mechanical design, cryostat design, and quench simulation. A novel coil shape is designed with the current being distributed on the racetrack shell. The magnetic field design results meet the requirements of the CEE spectrometer. The nominal operating current is 1580 A. The calculated displacements and stresses of the cold mass indicate that the structural design is reasonable and robust. For the cool-down of the cold mass, it combines both liquid helium and conduction cooling. The coil cases are closely fastened beneath the liquid helium vessels. Moreover, cooper strips are positioned between the coil cases and helium vessels to ensure the temperature uniformity. The calculated maximum temperature on the superconducting coils is 4.6 K which meets the operational requirement.

**Acknowledgments**

This work is supported by the National Natural Science Foundation of China under Grant No. 11927901.